\documentclass[12pt,english]{article}

\usepackage{geometry}
\usepackage[titletoc,title]{appendix}

\usepackage{ebgaramond-maths}
\usepackage{amsfonts, amsmath, newpxmath}

\usepackage{setspace}
\raggedright

\usepackage{sectsty}
\allsectionsfont{\normalsize\raggedright\centering}

\usepackage[]{tocloft}
\addtocontents{toc}{\cftpagenumbersoff{section}}
\addtocontents{toc}{\cftpagenumbersoff{subsection}}

\makeatletter
\renewcommand{\@cftmaketoctitle}{}
\makeatother

\usepackage[]{footmisc}

\usepackage{natbib}
\usepackage{url}
\makeatletter
\def\url@leostyle{%
    \def\UrlFont{\sf}}{\def\UrlFont{\small\ttfamily}}
\makeatother
\bibpunct
{(} 
{)} 
{,} 
{a} 
{} 
{;} 

\usepackage{authblk}

\usepackage{graphicx}
\usepackage{babel}
\usepackage{epigraph}

\usepackage{amsmath}
\usepackage{amsfonts}

\usepackage{framed}

\usepackage{xifthen}
\usepackage{cleveref}


\numberwithin{equation}{section}

\newcommand{\citetbjps}[2][]{\ifthenelse{\equal{#1}{}}{\citeauthor{#2} ([\citeyear{#2}])}{\citeauthor{#2} ([\citeyear{#2}], #1)}}
\newcommand{\citealtbjps}[2][]{\ifthenelse{\equal{#1}{}}{\citeauthor{#2} [\citeyear{#2}]}{\citeauthor{#2} [\citeyear{#2}], #1}}
\newcommand{\citepbjps}[2][]{\ifthenelse{\equal{#1}{}}{(\citeauthor{#2} [\citeyear{#2}])}{(\citeauthor{#2} [\citeyear{#2}], #1)}}
\newcommand{\citeyearbjps}[2][]{\ifthenelse{\equal{#1}{}}{[\citeyear{#2}]}{[\citeyear{#2}], #1}}
\newcommand{\citeyearparbjps}[2][]{\ifthenelse{\equal{#1}{}}{([\citeyear{#2}])}{([\citeyear{#2}], #1)}}
\newcommand{\citeposbjps}[2][]{\ifthenelse{\equal{#1}{}}{\citeauthor{#2}'s ([\citeyear{#2}])}{\citeauthor{#2}'s ([\citeyear{#2}], #1)}}

\newcommand{\sprsc}[1]{\ensuremath{^{#1}}}

\begin{document}

\title{{\Large {\sc Analog Computation and Representation}}}
\author{{\large {Corey J. Maley}}\\{forthcoming in \emph{The British Journal for the Philosophy of Science}}}
\date{}

\maketitle

\thispagestyle{empty}

\begin{abstract}
	Relative to digital computation, analog computation has been neglected in the philosophical literature. To the extent that attention has been paid to analog computation, it has been misunderstood. The received view---that analog computation has to do essentially with continuity---is simply wrong, as shown by careful attention to historical examples of discontinuous, discrete analog computers. Instead of the received view, I develop an account of analog computation in terms of a particular type of analog representation that allows for discontinuity. This account thus characterizes all types of analog computation, whether continuous or discrete. Furthermore, the structure of this account can be generalized to other types of computation: analog computation essentially involves analog representation, whereas digital computation essentially involves digital representation. Besides being a necessary component of a complete philosophical understanding of computation in general, understanding analog computation is important for computational explanation in contemporary neuroscience and cognitive science.

\end{abstract}

\epigraph{Those damn digital computers!}{Vannevar Bush, MIT}

\pagebreak

\mbox{} \\

\tableofcontents

\mbox{} \\

\section{Introduction}

Like clocks and audio recordings, computation comes in both digital and analog varieties. Relative to digital computation, analog computation has been neglected, and as a result, not well understood. This is partially due to the fact that an account of what counts as analog in general has proven controversial, particularly in contrast to what counts as digital. Fifty years have passed since \citetbjps{goodman1976languages} began the discussion of the so-called analog/digital distinction in the philosophical literature, and still we have not reached a consensus. In the few places where analog computation has specifically been mentioned, it usually goes something like this: `Analog computation is often contrasted with digital computation, but analog computation is a vague and slippery concept\ldots Roughly, abstract analog computers are systems that manipulate continuous variables to solve certain systems of differential equations' \citepbjps[p. 123]{Piccinini:2015ut}.

One might think, like Piccinini, that all we need to know about analog computation is that it is continuous, rather than discrete. One might think that we need no more than this kind of rough characterization of analog computation because analog computation is no more than a historical curiosity, devoid of interest or relevance to contemporary philosophy. But these thoughts would be misguided.

The received view of analog computation---that it is essentially about continuity---is simply wrong, as shown by studying actual analog computers from the 20\sprsc{th} century. Providing an account of analog computation is not as simple as the received view would have it. Further, while it is true that the heyday of analog computers has come and gone (there are no companies that produce analog computers anymore), there are two reasons why providing such an account is still important. First, if we want to understand computation simpliciter, we need a clear account of analog computation in order to see how it might fit into a more general account of computation that includes analog and digital (and perhaps other) types as species. Second, if we take seriously the idea that cognitive science and neuroscience are in the business of explaining what minds and brains do in terms of the computations that they literally perform,\footnote{This is a stronger claim than that the brain can be computationally simulated, which is true of virtually any scientific object of interest \citepbjps{Piccinini:2007we}.} then we should understand all types of computation that might be applicable to such explanations.

Here is the structure of what follows. In order to get a feel for how analog computation actually works, I will first present a few examples of different types of mid-20\sprsc{th} century analog computers. After making clear why continuity alone does not suffice, I will then argue that a refined version of analog representation---the Lewis-Maley\footnote{This is how \citetbjps{Adams:2019wo} labels the view originally developed by \citetbjps{Lewis:1971ug} and refined and defended by \citetbjps{Maley:2011jv}.} view---explains what is analog about analog computation. Finally, with this account of analog computation in hand, I will conclude with some general considerations about the relevance of this account to contemporary issues in philosophy and the cognitive sciences.

Before beginning in earnest, a preemptory note is in order. Space prohibits a full account of analog and digital computation. As just mentioned, the standard approach is to present an account of both simultaneously. I think this has been a mistake, because, `analog' and `digital' are neither opposites not jointly exhaustive. I can only offer some comments on digital computation and representation in passing. It should simply be noted that a system of representation or computation that is not analog on the account presented here is not necessarily digital.

\section{Analog Computers}
\label{sec:analog_computation}

If one wants to trace the history of digital computers, one excellent place to start is the work of \citetbjps{Turing:1938ui}. To be sure, digital computing machines predate Turing's work: examples include Babbage's difference engine and Pascal's calculator (sometimes known as the Pascaline). Nevertheless, Turing's work on computable numbers is usually taken to have initiated what we now understand as computer science, and later, digital computer engineering.\footnote{However, some, such as \citetbjps{Corry:2017ir}, have argued that Turing's theoretical work did not influence the birth of digital computation nearly as much as is commonly thought.}

There is no such conceptual birthplace when it comes to analog computers. Analog devices have been used since antiquity: an early review essay on analog computing machines begins `The use of instruments of computation and analysis is as old as mathematics itself' \citepbjps[p. 649]{Bush:1936ve}, and the function of the two-millenium-old Antikythera mechanism has only recently been discovered \citepbjps{Efstathiou:2018ex}. As for a contemporary theoretical centerpiece---something like the `Turing' of analog computation---common citations include \citepbjps{Shannon:1941tn} and \citepbjps{PourEl:1974te}; but, as we will see later, even these are insufficient to account for analog computation in its entirety.\footnote{For instance, a recent monograph \citepbjps{Piccinini:2015ut} on physical computation states that `[t]he clearest notion of analog computation is that of Pour-El (1974)'.}

Rather than tracing a complete history of these instruments, we will begin with those devices first developed in the 1930s, now referred to as the first analog computers.\footnote{Using the term `computer' to describe these machines is, of course, anachronistic: this term referred to a particular job, usually held by women, that involved performing mathematical computations by hand. More accurate would be `analog computing machines.' For brevity, however, I will use the anachronistic terminology in what follows.} Even still, beginning here is something of a challenge because of the lack of attention paid to these machines, particularly relative to digital computers. One historian puts the point this way:
\begin{quote}
In general, historians of computing have neglected analog computing, viewing it primarily as an obsolete predecessor to digital\ldots [W]e have not yet begun to understand the history and significance of analog computing, especially the relationship between analog and digital machines. \citepbjps[p. 10]{Mindell:2002vm}
\end{quote}

A computer scientist makes a similar point:
\begin{quote}
Because digital computers and computation have been so successful, they have influenced how we think about both computers as machines and computation as a process---so much so, it is difficult today to reconstruct what analog computing was all about. \citepbjps[p.3]{Nyce:1996ue}	
\end{quote}

To be sure, there have been some limited discussions of analog computation in the philosophical literature. For example, O'Brien and Opie (\citeyearbjps{OBrien:2008df}, \citeyearbjps{OBrien:2010vn}) use analog computation to illustrate the role of representation in cognitive science. Shagrir (\citeyearbjps{Shagrir:2010dm}, \citeyearbjps{Shagrir:tx}) discusses the use of what he calls `analog-model' computing, which he takes to involve the simulation or modeling of a system.\footnote{Shagrir's view has much in common with the view on offer here, although his focus is on computation more generally. The differences are important, but they will have to wait for another time.} \citetbjps{Isaac:2018ke} has recently argued that analog computation is compatible with embodied approaches to cognition. \citetbjps{Papayannopoulos:2020ik} has recently defended the orthodox view that analog computers are simply computers with continuous values (a point I will return to below). These and related works are undoubtedly important steps toward the project of understanding analog computation; nevertheless, we do not yet have a complete philosophical treatment of the subject. Providing such a treatment is one goal of this essay.

I will describe three somewhat simplified examples of analog computers: one mechanical, one electrical, and one electromechanical. These examples will illustrate some of the general principles needed to motivate later discussion of the right way to think about analog computation.

\subsection{Mechanical analog computers} 
\label{sub:mechanical_analog_computers}
\begin{figure}[!htb]
	\begin{center}	
	\includegraphics[width=5in]{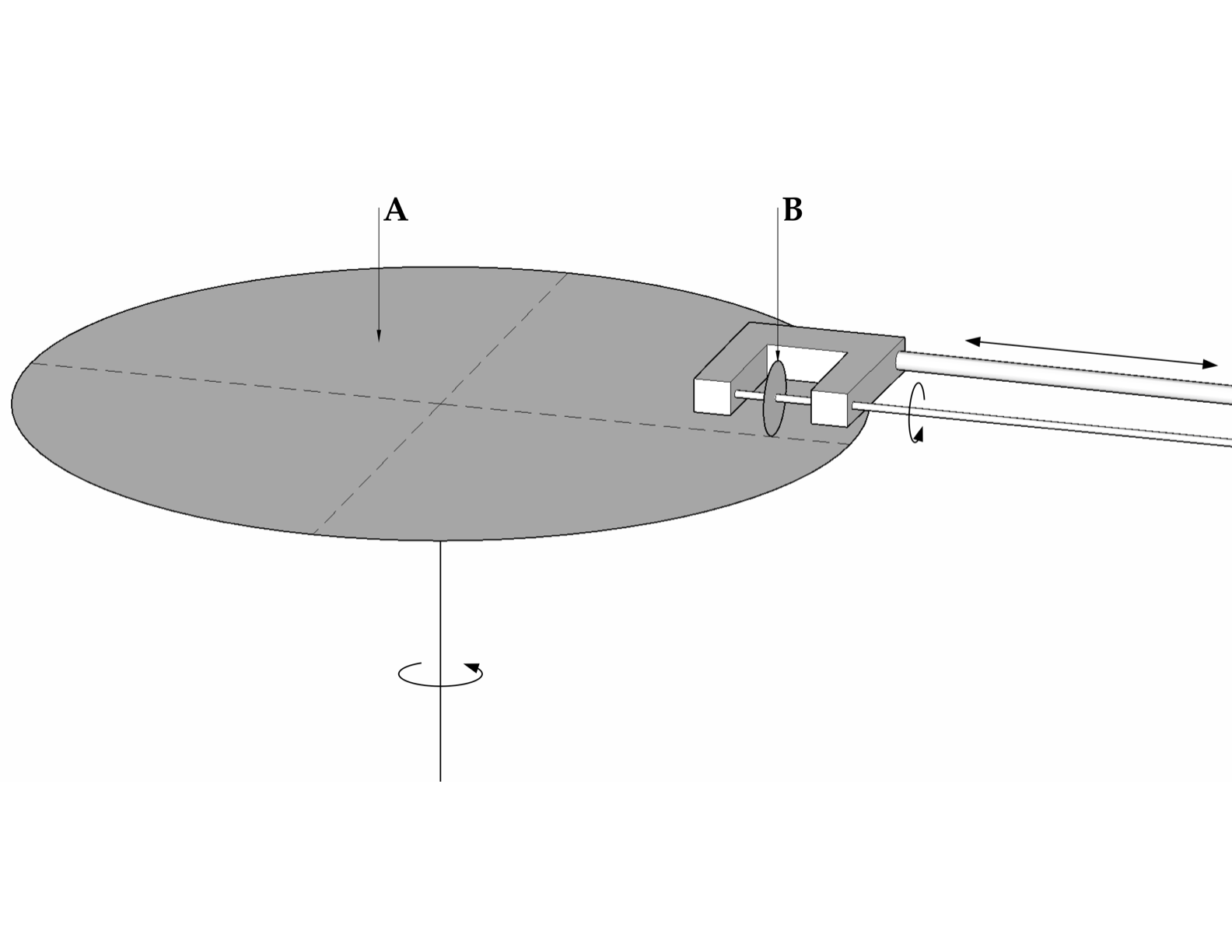}
	\caption{\emph{Disk B rotates faster near the edge of disk A than near the center.}}
	\label{fig:disk}
	\end{center}
\end{figure}

One of the most well-known mechanical analog computers is the differential analyzer, developed by Vannevar \citetbjps{Bush:1931cb}. `Mechanical' in this context means that the elements doing the computing use physical movement, such as rotation and displacement, to perform their tasks. Bush's differential analyzer uses an interconnected series of components to compute solutions to mathematical operations, including differential equations. One such component is the integrator, which represents the values of functions (changing over time) using the rotation of rods connected to a disk and ball bearing assembly. A simplified version of this component\footnote{Simplifications include using two connected disks rather than a disk and ball bearing assembly, as well as omitting various supporting structures to allow for easier visualization.} is shown in Figure \ref{fig:disk}. First, we will examine how it works mechanically, then how it functions as a computer. 

There are two disks: the large bottom disk A, which is a turntable, and disk B, in contact with, and perpendicular to, disk A. Disk A rotates at a constant speed, and disk B is connected to an input linkage that can be slid left and right (closer to and farther from the center of A), changing the point where B contacts A. As A rotates, B rotates in the perpendicular direction. However, while the speed of A is constant, the speed of B depends on exactly where B contacts A: disk B will rotate faster when near the edge of A, slower near the center of A, and not at all when it is at the exact center of A.

So how, and what, does this mechanism compute? As the name implies, this component integrates a function over time.\footnote{Strictly speaking, the independent variable could be something other than time, but time is the most common and easiest to explain, so that is what we will use.} The value of the function determines the position of B, which moves to the right (farther from the center) as the value increases, and left (close to the center) as that value decreases.

\begin{figure}[!htb]
	\begin{center}	
	\includegraphics[width=5in]{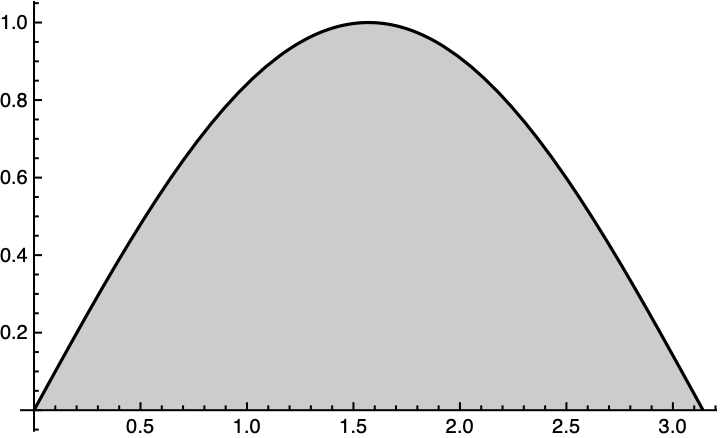}
	\caption{\emph{The sine function from zero to $\pi$.}}
	\label{fig:sine}
	\end{center}
\end{figure}

As an example, suppose we want to integrate part of the sine function shown in Figure \ref{fig:sine}. This means we want to compute the area under the black curve (shown in gray). The bottom disk A serves as the independent variable. Because the function we are integrating starts at zero, the position of B would begin at the center of A (where, again, even when A is moving, B will not move in response). As our function begins to increase, the position of B will move to the right, further from the center. As it does so, B will start rotating, and rotate faster as it moves closer to the edge of A. At the apex of the function, B will be closest to the edge of A; after that, B will move back toward the center of A. By tracking the total number of rotations of B, we compute the area under the curve.

In summary, the rotation of disk A represents an independent variable, such as time. Once the computation begins, A rotates at a constant speed, not influenced by other variables. The position of B, relative to the center of A, serves as the input, or dependent variable. This position represents the value of the function to be integrated. For different functions, this position will move back and forth in different ways: in the case of the sine function, it will decelerate as it moves toward the center, then accelerate away from it after reaching the maximum of the function.\footnote{Note that the disk can represent both positive and negative values of a function: in this example, if the disk were to move to the left of the center of A, then it would subtract the total number of rotations, just as we would expect when integrating negative values of a function.} Again, the speed that disk B rotates is determined by where it is relative to the center of A; thus, the running total number of rotations of the disk B serves as the output: it represents the definite integral---the area under the curve---of the input up to that point.

Many other devices were used to compute various mathematical functions in similar ways: displacement and rotation were used to represent quantities, which were then mechanically engaged in clever ways to deliver the requisite output. The particular example of the integrator is merely one component, which could be connected to other components to compute more complex functions. The output of the integrator, for example, might serve as the input to another component (and even more integrators in the case of higher-order differential equations). Similarly, the input could come from the output of a separate component (in fact, the next section will demonstrate an example where a number of elements are connected together in just such a way). Besides the integrator, there were mechanical adders, multipliers, function generators, and many others (presented in, for example, \citepbjps{Soroka:1954wj}, \citepbjps{Truitt:1960wr}, and \citepbjps{Ashley:1963uv}).

\subsection{Electronic analog computers} 
\label{sub:electronic_analog_computers}
Rather than using moving mechanical parts, electronic analog computers largely consisted of configurable electronic circuitry. Thus, in these machines, quantities are represented by various electrical properties (such as voltage or resistance) rather than mechanical properties. Once again, we will look at a simplified example in order to give us a feel for how these machines worked.\footnote{This example is adapted from a textbook on analog computers \citepbjps[p. 29]{Peterson:1967uw}.}

\begin{figure}[!htb]
	\begin{center}	
	\includegraphics[width=4.0in]{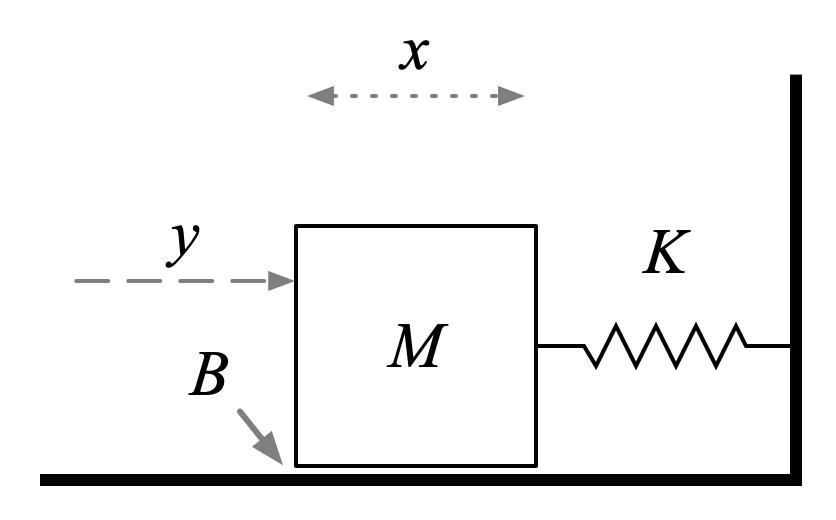}
	\caption{\emph{A mechanical system consisting of a mass M connected to a wall via a spring (with spring-constant K), friction (B), with a force applied to its left side (y).}}
	\label{fig:MechanicalDiagram}
	\end{center}
\end{figure}

First, let us consider a simple physical problem that we can set up an analog computer to solve. In Figure \ref{fig:MechanicalDiagram}, we have a mass M connected to a wall via a spring. The problem we would like to solve is: how does the mass move as a force is applied to its side? In other words, what is its left-right displacement (the variable $x$) when a right-moving force (the variable $y$) is applied to the left side of M and released?

This problem is not trivial: we must take into account the friction between the mass and the ground (given by B), and the `springiness' of the spring (given by K). Still, we can use some basic physics to characterize this system mathematically. Using the values M = 1 kg, B = 3 nt/m/sec, K = 16 nt/m, and $y$ = -80 nt, the differential equation (and initial conditions) for this system is given by the following:
\begin{align} \label{eq:MechanicalEquations}
	& y = x'' + 3x' + 16x;\ y = - 80 \\
	& x'(0) = -0.64;\ x(0) = 2 \nonumber
\end{align}

To set up an electronic analog computer to provide the solution, we will rewrite the equation to isolate the second derivative, which gives us:
\begin{equation} \label{eq:oneline}
	 -x'' = 3x' + 16x - y
\end{equation}

\begin{figure}[!htb]
	\begin{center}	
	\includegraphics[width=4in]{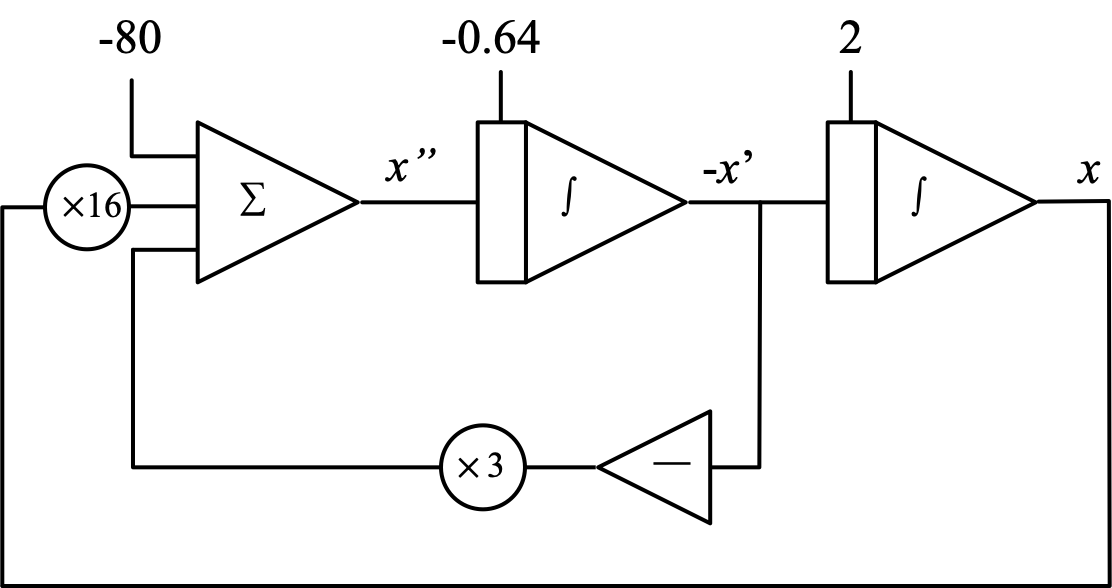}
	\caption{\emph{Electronic analog computer schematic.}}
	\label{fig:MechanicalAnalogComputer}
	\end{center}
\end{figure}

Using this equation and the initial conditions as a guide, we connect four types of electronic components together, as shown in Figure \ref{fig:MechanicalAnalogComputer}: an adder (triangle with $\sum$), two integrators (triangles with $\int$), an inverter (triangle with $-$), and two potentiometers (circles with $\times$), which act as multipliers.

\begin{figure}[!htb]
	\begin{center}	
	\includegraphics[width=4in]{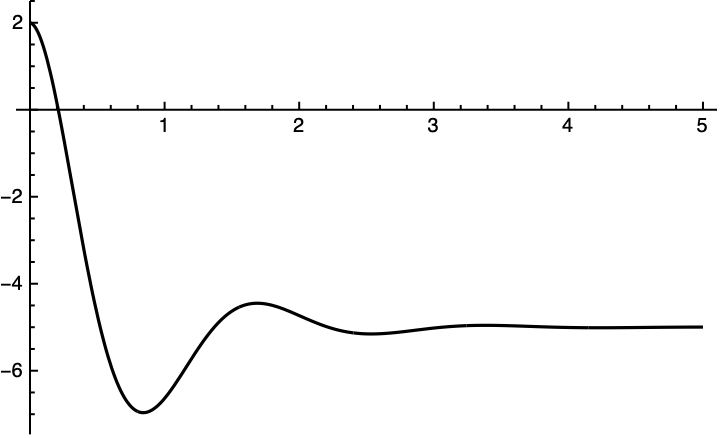}
	\caption{\emph{Solution of the system given in Equation \ref{eq:MechanicalEquations}.}}
	\label{fig:MechanicalSolution}
	\end{center}
\end{figure}

When the voltages---which are determined by the values of the equations---are `run' on this computer, the output of the system is given by $x$. The output starts at a value of two volts (the initial value given in the equations), then changes over time as a function of itself (note that the output $x$ is fed back in as an input to the computer in Figure \ref{fig:MechanicalAnalogComputer}) and the other terms in the equation.\footnote{In real electronic analog computers, the values of the variables of interest would have to be scaled so as to stay within the electrical limits of the analog computer. So, for example, if our system required a value of 700, we might scale everything by .01 so that the computer would use seven volts, rather than 700 volts. In such a case, the graphical output would look the same, but the axes would need to be linearly scaled to make sure the values are correct. This is identical to how slide rules can only be used within a certain range, so problems have to be scaled up (or down) in order to fit in that range, then scaled back down (or up) once the calculations are finished.} After a rapid change and a few oscillations, the output reaches a steady state, as shown in Figure \ref{fig:MechanicalSolution}. This is what happens with the physical system depicted in Figure \ref{fig:MechanicalDiagram}: the mass rapidly goes back to---and then overshoots---its equilibrium point, oscillates a bit around that point, and finally settles.

This particular type of analog computer is known as an electronic differential analyzer (EDA). EDAs were by far the most common type of electronic analog computer, simply because differential equations are so common in science and engineering (most people who are familiar with analog computers generally have in mind an EDA). We need not attend to the details of the individual components of the EDA, but we can see how the overall structure of this computer solves the given equations. For example, consider the adder on the top left (indicated by the triangle with the summation sign), which sums its three inputs. The first (top) input to that component is simply the value -80, which here is represented by -80 volts. Tracing the path of the other two inputs, we see that the bottom one is the output from the first integrator (which is $x'$), multiplied by -3. The middle input is the result of the output from the second integrator (which is $x$), multiplied by 16. This is exactly what Equation \ref{eq:MechanicalEquations} specifies.

The design and physical implementation of the electronic analog computer looks very different than the mechanical one illustrated earlier. The important point to note, however, is that the basic principle is the same. In each case, we have a physical quantity representing a variable: rotation and displacement in the mechanical case, and voltage in the electronic case. Both computers are designed so that they manipulate the physical quantities that represent values in ways that correspond to the requisite mathematical operation.

Insofar as the relevant physical quantities are continuous, both the mechanical and the electronic analog computers illustrated here fit perfectly well with the received view of analog computation mentioned at the beginning of this essay: analog computation is simply computation using continuous elements. However, analog computers also used discontinuous elements, an important but little-known fact that causes serious trouble for the received view. So let us look at some examples of these discontinuous analog computers. 

\subsection{Discontinuous analog elements} 
\label{sub:discontinuous_analog_elements}
While many analog computer elements are continuous, including the EDA mentioned above, not all are. One textbook introduces this point quite nicely:
\begin{quote}
	Ninety-nine and forty-four one-hundredths percent of the time, when an engineer speaks of an analog computer he is referring to an \emph{electronic differential analyzer} (EDA), but the EDA is just one type of analog computer, one specific application of the general principal of computation by analogy. So let's see first of all what is meant by \emph{analogy} and how we use analogs in computation---in the general sense, not just in the EDA. \citepbjps[p. 1, emphasis original]{Peterson:1967uw}
\end{quote}
Let us follow this lead and look at that 0.56 percent.

Many phenomena that one might want to study using analog computers include discontinuities of various sorts; as such, analog computers implement many different kinds of discontinuities.\footnote{For the moment, what is meant by `discontinuous' is that the function in question is not smooth, which in turn means that it has a number of points at which the function is not differentiable. Although this accords with a commonsense understanding of discontinuous, it would still count as continuous in the strict mathematical sense, in which the function has gaps or jumps. Later we will see examples of functions that are discontinuous even in this strict sense.} For example, some physical systems that involve spur gears or other mechanical parts have a certain amount of slack that cannot be eliminated, because the components cannot be be in perfect physical contact (if they were, they would be unable to move). In the case of gears, this means that one gear might begin moving for a very short time before it contacts another, at which point the second gear will move. Thus, the movement of the second gear is not a continuous function of the movement of the first. In fact, if several such gears are connected, the slack can become an important feature of the system to be studied. As such, it was necessary to include this kind of discontinuity in analog computers to study systems with such discontinuities.

\begin{figure}[!htb]
	\begin{center}	
	\includegraphics[width=5in]{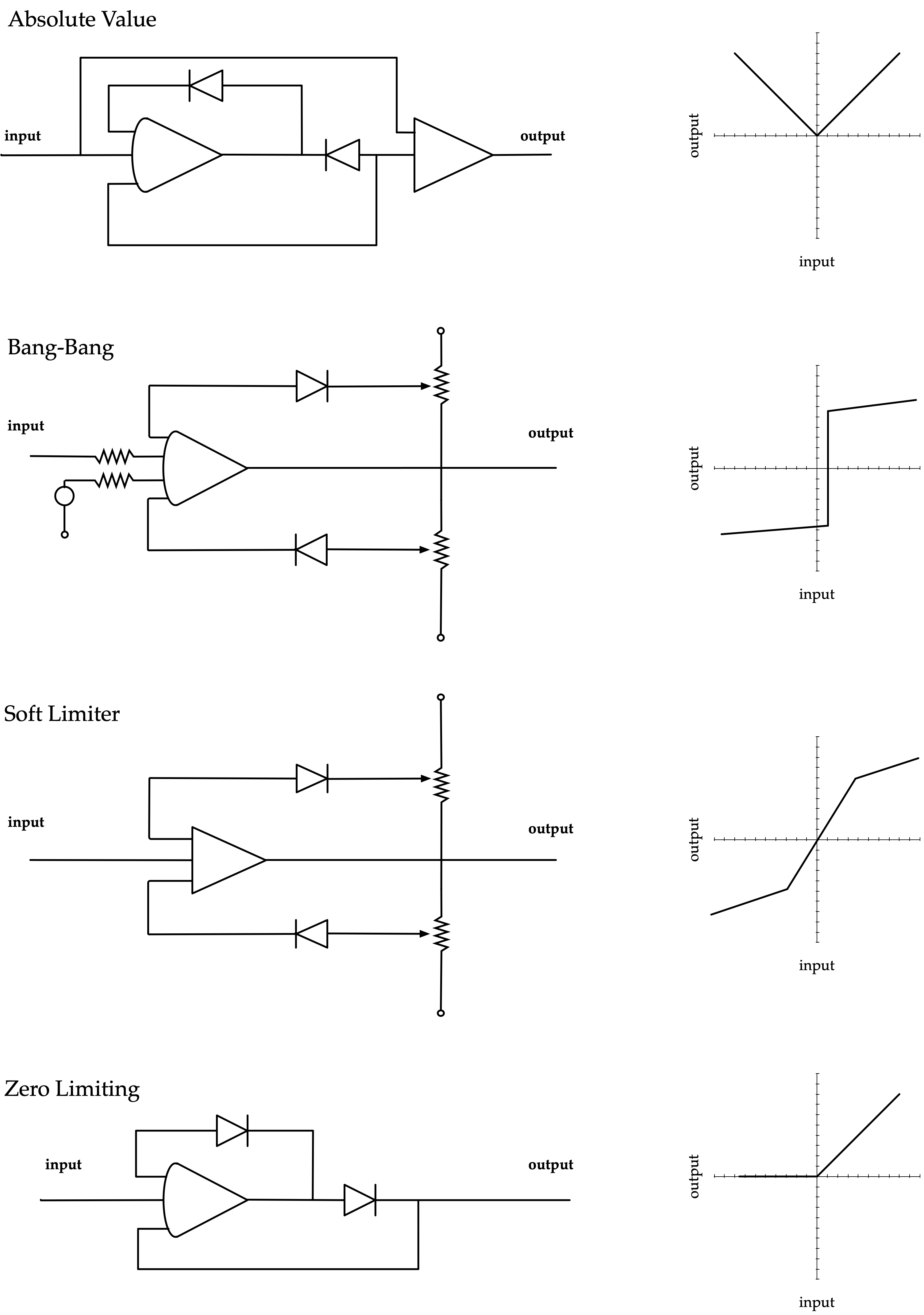}
	\caption{\emph{Four different circuits implementing discontinuities (left column) and their respective graphs (right column).}}
	\label{fig:fourCircuits}
	\end{center}
\end{figure}

More generally, we may want to model any number of discontinuous functions, for any number of reasons. Figure \ref{fig:fourCircuits} illustrates a few of these as they are implemented in an electronic analog computer, adapted from \citepbjps[p. 31]{Cadman:1969tg}. Note that the behavior of the second gear just mentioned is captured by using the `zero limiting' circuit: as a function of the first gear's movement, the second gear is zero until a single point, at which it abruptly (not smoothly) begins to increase.

A more complex example is also one of the more surprising, particularly to those who endorse the received view of the analog, and who are only familiar with the EDA as the paradigm example of analog computation. Recall the example of the spring-mass system and the accompanying EDA from \cref{sub:electronic_analog_computers}. We began with a physical problem, which we were then able to precisely characterize in mathematical terms. This allowed us to create a circuit based solely on that mathematical characterization, which in turn allows us to compute the solution via an EDA. In this case, this was because we could analyze the problem using known physical principles: characterizing spring-mass systems in terms of differential equations is a well-known technique.

\begin{figure}[!htb]
	\begin{center}	
	\includegraphics[width=4in]{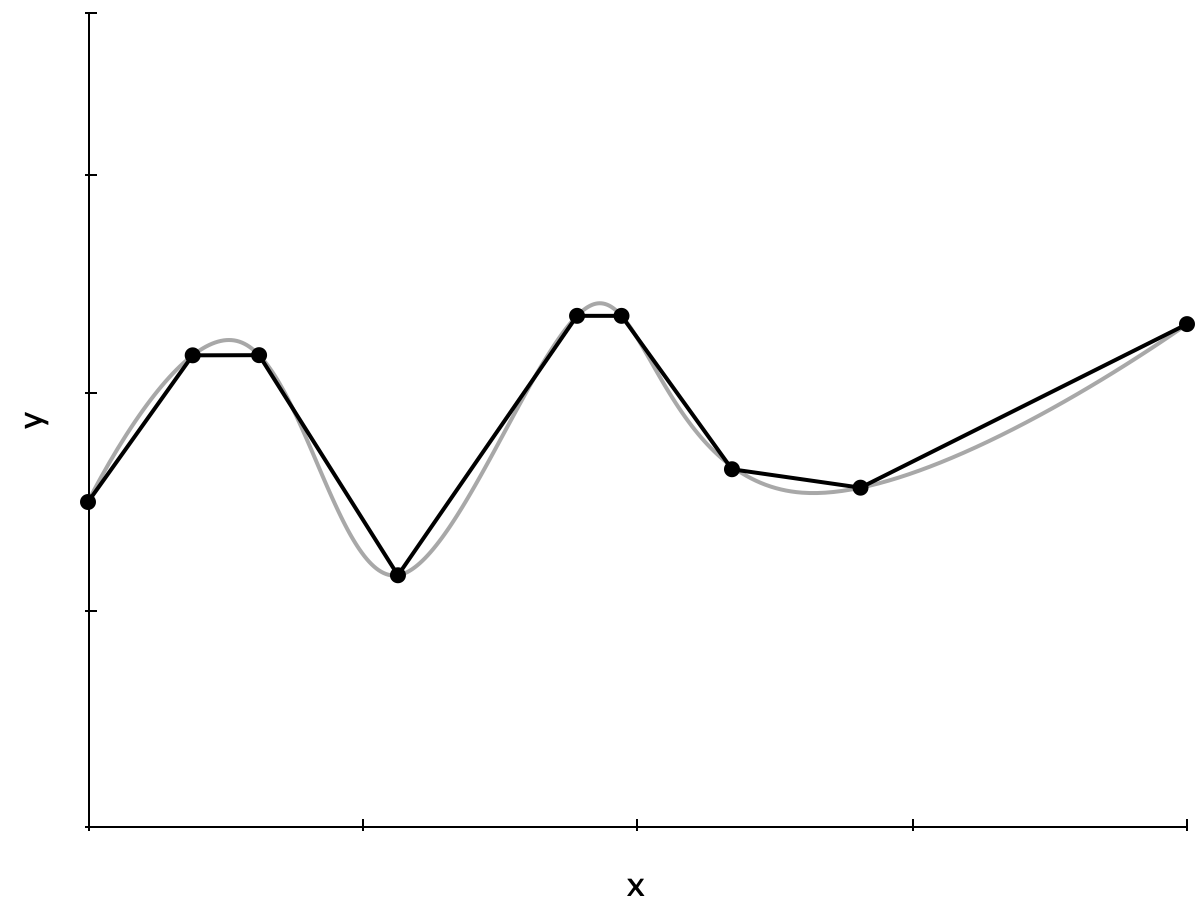}
	\caption{\emph{A continuous function (grey), approximated by a piecewise-linear function (black).}}
	\label{fig:pointwise}
	\end{center}
\end{figure}

Some problems, however, do not admit of this kind of mathematical characterization. For example, we may know what a particular function looks like, although we do not know how to translate that into a set of equations that we can then use to construct an analog computer. Many problems are unlike the spring-mass system in that regard: there may be no first principles from which one can derive a set of equations that describe the system. Although this might seem to render analog computers useless for such problems, they have components to handle cases exactly like these:
\begin{quote}
	Such behavior presents almost insurmountable obstacles to purely mathematical investigation, but poses no particular difficulty to analog-computer investigation. Again, we are not solving equations, we are modeling systems. Thus if we can describe the input-output relationship\ldots, all we need to do is provide an element on the computer which has the same relationship between its input and output voltages. Such elements are known as \emph{arbitrary function generators.} \citepbjps[p. 109, emphasis original]{Peterson:1967uw}
\end{quote}
Using an arbitrary function generator, an analog computer could be set up to construct a discrete, piecewise-linear approximation to any function, even one without a known mathematical characterization. The piecewise-linear approximation generated by this component consists of a series of straight line segments, with discontinuities where those segments meet. Figure~\ref{fig:pointwise} shows an example of a continuous function plotted with such a piecewise approximation. Depending on the application and the particular function generator used, better approximations could be achieved by varying the number of points and the distances between the points.

If---as the received view would have it---analog computers essentially use only continuous elements, one would expect that an analog computer would have to use a continuous function to approximate a discontinuous problem of interest. But note that the exact opposite is happening in the example just given. For some applications in which one wanted to study a continuous function whose mathematical characterization is unknown, the analog computer could use a discontinuous function to model that continuous function. 

In an earlier footnote, I mentioned that `discontinuous' in the examples just given simply means that the functions in question are not smooth. Although analog computer users and engineers referred to these function as discontinuous, mathematically speaking these kinds of function are still continuous. The rough idea is that if we were to draw these functions on a piece of paper, we would not need to lift our pencil off of the surface, even if there are sharp (rather than smooth) changes in direction.\footnote{More technically, these functions are not everywhere-differentiable.} Mathematically discontinuous functions require gaps or jumps where we would need to lift our pencil off of the surface to draw them. As it turns out, analog computers implemented these types of functions, too.

\begin{figure}[!htb]
	\begin{center}	
	\includegraphics[width=5in]{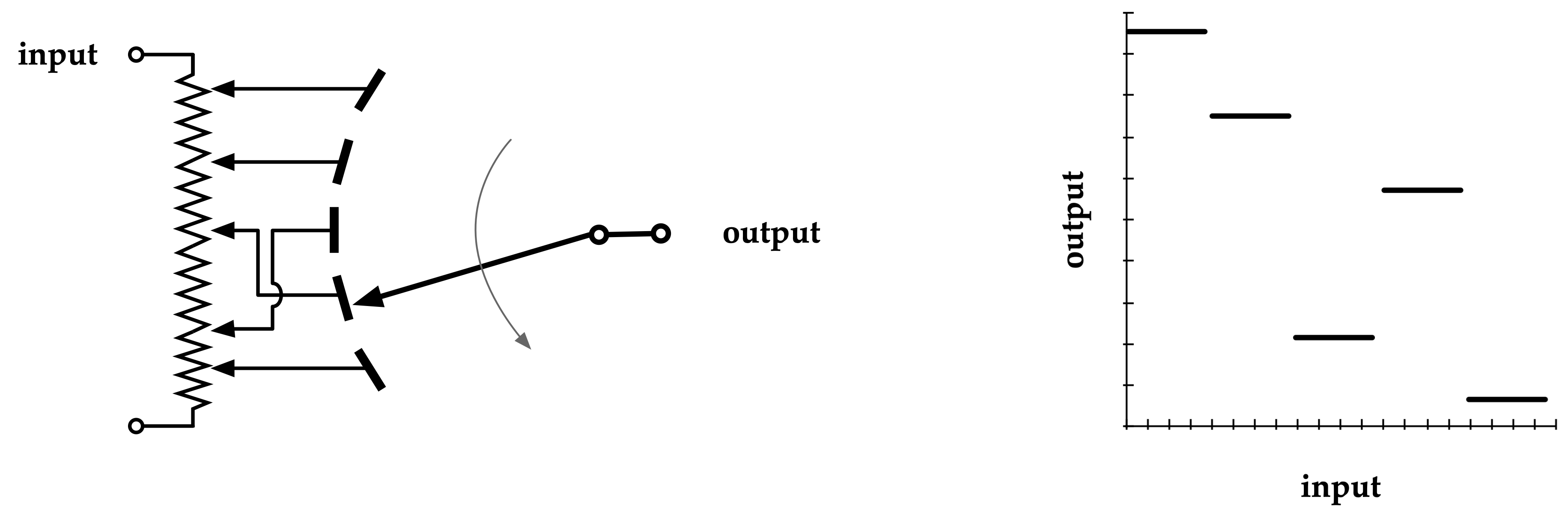}
	\caption{\emph{Switch-type step function generator (left) and its resulting step function (right).}}
	\label{fig:commutator}
	\end{center}
\end{figure}

An excellent example is a particular electromechanical component. Like purely electronic analog computers, electromechanical analog computers use electrical properties, such as voltage, to represent variables of interest. However, in the electromechanical case, these variables are manipulated by mechanical means rather than purely electronic means. The example here is a step-function generator. Step-functions are constant for a specified interval, then `jump' to a different constant for a different interval. Figure \ref{fig:commutator} shows a schematic of a simplified version of such a component, plus the step function it generates (adapted from \citepbjps[p. 254]{Korn:1952wy}). The component has a rotating element that makes contact with separate wires, which in turn connect to a variable resistor at different points. The farther from the input, the lower the resistance, and thus the greater the output. As this element rotates (counterclockwise in this example), it momentarily breaks contact with one wire, then makes contact with the next wire, resulting in a discontinuous jump from one voltage to another.\footnote{This particular example is of a non-shorting switch, or `break-before-make' contact. This ensures that the voltage truly jumps from one value to another. Other switches had overlapping contacts, called shorting switches, or `make-before-break' contact, which could be used if the true discontinuity was not wanted. } By adjusting the number of wires, where they contact the resistor, and the speed of the rotating element, one could implement different step functions with different characteristics.

Once again, we have a counterexample to the received view. These elements (and ones like them) were not uncommon in analog computers, yet they do not use continuously-varying elements. Instead, the variation is as discontinuous as could be.

\begin{figure}[!htb]
	\begin{center}	
	\includegraphics[width=5in]{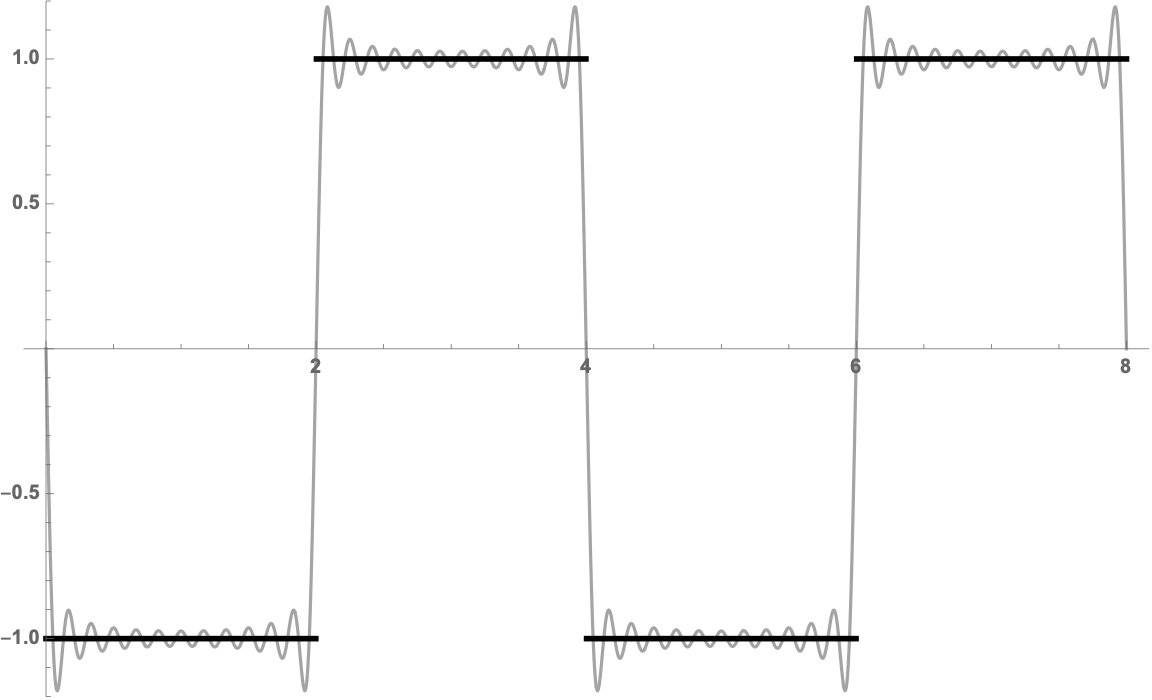}
	\caption{\emph{Step function (black) and a Fourier series continuous approximation (grey).}}
	\label{fig:stepgibbs}
	\end{center}
\end{figure}

At this point, one might wonder why analog computer engineers would go to the trouble of making these discontinuous components, given the well-known fact that these types of functions can be approximated by continuous functions. For example, a simple step function that goes back and forth between two values can be approximated by a sum of sine and cosine functions. While this may be good enough for some purposes, it is not good enough for others. In particular, the difference between the approximation and the actual value of the step function one wants can be particularly bad at transition points for certain kinds of approximations,\footnote{This is known as the Gibbs phenomenon.} as illustrated in Figure \ref{fig:stepgibbs}.
So, rather than trying to use continuous functions to approximate discontinuous functions, analog computers were able to directly implement discontinuous functions. In a discussion of the use of relays and switches to incorporate a discontinuous voltage change (like the electromechanical component just illustrated), rather than using a continuous approximation (the diode limited amplifier circuit, which would result in an approximation similar to that shown in Figure \ref{fig:stepgibbs}), the author of an analog computing monograph explains:
\begin{quote}
	For the problem being studied, it is not immediately obvious why the relay is needed. The voltage from the diode limited amplifier circuit can be made to closely approximate a delayed step function.\ldots There are two reasons why this is not practical. The slope of the `step' function out of this circuit is not exactly zero after the original discontinuity. In the process of adding two of these step functions, a small error that increases with [time] would be applied to the integrator and cause an unwanted `drift' in the output of the integrator. \citepbjps[p. 201]{Ashley:1963uv}
\end{quote}
The idea is simply that small approximation errors in different components can combine to form increasingly large errors. Better to use the exact values of a discontinuous function to model a discontinuous function, rather than a continuous approximation.

These are just a few of the examples that demonstrate how analog computation was not only about continuity. Discontinuous components are important for approximating continuous functions, as well as their use in directly implementing discontinuous functions of interest. Although this may seem surprising today, it was known during the heyday of analog computation. For example:
\begin{quote}
	Any mechanism which involves continuous variables is nowadays in danger of being called ``analog," of course to distinguish it from ``digital"\ldots To make matters still more confounded, the common usage for computing structures, whereby only continuous methods are called analog, is wrong, since it is clear that discrete or digital machines may also embody and constitute analogs of prototype phenomena. \citepbjps[p.7]{Philbrick:1961te}
\end{quote}

It is clear that the received view---that analog computation is essentially about continuity---is simply false. However, we still need to make clear what analog computation is essentially about. In the next section, I will discuss different accounts of the analog, and argue that only one particular account of analog representation makes sense of analog computation.

\section{What makes analog computation `analog' and `computational'} 
\label{sec:accounts_of_analog_representation}
Nearly all accounts of the analog focus on analog representation, rather than analog computation more specifically. Moreover, nearly all accounts of analog representation are made in contrast with digital representation. This is unfortunate. The similarities and differences between analog and digital representation---and analog and digital computation---are lost when `analog' and `digital' are taken to be opposites, and jointly exhaustive of representational types. In other words, many have assumed that once we know the right way to characterize analog representation, then digital representation comes for free (or vice versa): digital is whatever is not analog. While I take this assumption to be false, I cannot engage in a full defense here, so I will limit the discussion to the relevant aspects of the positive accounts of what counts as analog. I will first review several accounts of analog representation, and then argue that only the so-called Lewis-Maley account is able to make sense of analog computation. What follows is captured quite well by Peacocke: `Analog representation is representation of magnitudes, by magnitudes. Analog computation is the operation on representing magnitudes to generate further representing magnitudes' \citepbjps[p.52]{Peacocke:2019vm}, although how that slogan is elaborated here differs from Peacocke's own elaboration. Again, I will not provide a detailed account of digital representation (much less digital computation), although I will offer a few remarks to make clear the contrasts that other authors have made, and outline the case for why `digital' is neither synonymous with `discrete', nor with `non-analog'.

Philosophical accounts of analog representation generally fall into one of two camps, which \citetbjps{Beck:2018kk} calls the `continuous' and the `mirroring' conceptions.\footnote{There are still other types, such as \citepbjps{Frigerio:2014kk}, that do not fit well into either camp; but these accounts are not relevant to the concerns noted here.} I will use  `covariation' instead of `mirroring' (for reasons explained below), but the idea is roughly the same. In what follows, we will look at each type of account, plus some specific instances of each.

\subsection{Analog as continuity} 
\label{sub:analog_as_continuous}
According to the continuous conception (the received view), the essential feature of analog representation is that it is continuous in nature. What exactly `continuous' means is not always made precise, but the basic idea is that analog representations vary smoothly, rather than in discrete steps.

The first account of this kind in the philosophical literature is due to \citetbjps{goodman1976languages}. On this view, analog representations are continuous, or dense, while digital representations are differentiated, or discrete.\footnote{Goodman speaks of representational schemes, rather than individual representations. The idea is that a single representation in isolation is neither continuous nor discrete, but representations can vary continuously or discretely in accordance with a scheme. While this is an important consideration, for brevity of exposition, I will simply refer to representations, rather than representational schemes.} \citetbjps{haugeland1981analog} draws from this account, and distinguishes between analog and digital devices in terms of the reliability of the procedures that read and write representation tokens. In short, digital devices read and write tokens that are completely determinate, with read/write procedures that are perfectly reliable. Analog devices, on the other hand, read and write tokens that are not perfectly determinate, with read/write procedures that are approximate at best.

\citetbjps{Katz:2016jc}, \citetbjps{Schonbein:2014dr}, and \citetbjps{Papayannopoulos:2020ik} all offer elaborations and defenses of these views. Katz clarifies a potential flaw in Haugeland's view, specifying that whether a given representation system is analog or digital is not a matter of objective facts about the system, but how the system is used (or supposed to be used). Katz makes clear that what counts as a user can be very general, and need not be a human or other agent located outside of the system in question. The point is simply that we need to look carefully at the context in which representations are read, written, and otherwise used. Schonbein argues for the received view on historical grounds. On this view, there is an entrenched engineering literature that treats analog as continuous, and digital as discrete. However, Schonbein also allows that there may be different varieties of analog representation, such that different accounts might be better suited for different purposes. Papayannopoulos argues that a modified version of Goodman's account is best suited for understanding analog computation, although he specifically discounts discrete components as being analog (although he does not use actual examples, he contends that two continuously-varying wheels connected together would be analog, but wheels that move in discrete steps, constrained by the teeth of gears, would be digital).

Given the discontinuous examples from \cref{sub:discontinuous_analog_elements}, this family of accounts of the analog does not properly characterize analog computation. There are, of course, differences among this family of accounts, and some of them disagree about hypothetical cases. However, what unites them is the thought that continuity is essential for any account of analog representation (or devices that use analog representations, such as computers). Given that each implicitly accepts the idea that analog and digital are both opposites and jointly exhaustive, each account takes discontinuous cases like the one presented above to be non-analog, and thus digital.

In summary, according to the continuous account, discontinuous analog computers are not really analog. That is a shortcoming: although not as widely known as the continuous elements of analog computers, discontinuous elements were not uncommon. Fortunately, another account of analog representation is available, which does properly characterize analog computation as analog.


\subsection{Analog as covariation} 
\label{sub:analog_as_covariation}

The second family of accounts of analog representation rejects the idea that this kind of representation is necessarily continuous. Instead, this family takes the essential feature to be some kind of mirroring, or covariation, between the representation and what is represented. What exactly `covariation' means is not always made precise (and differs somewhat between different accounts), but the basic idea is that analog representations are structurally isomorphic (or, in some cases, homomorphic) to what they represent.

The first account of this kind in the philosophical literature is due to \citetbjps{Lewis:1971ug}. Lewis suggests---contra Goodman---that what is important about the voltage in an analog computer is not that it is continuous, but that it covaries with what it represents. Specifically, as the quantity we want to represent increases, so does (for example) the voltage, which is what is doing the representing. The value 34, for example, is represented by 34 volts; 34.8 is represented by 34.8 volts. Furthermore, this covariation would occur even if we had discrete, rather than continuous, variations of voltages: whether the voltage could only increase in increments of 1, volt, 0.1 volts, or continuously, the covariation would still be present. According to Lewis, what does the representing in an analog representation is some primitive or `almost primitive' physical magnitude, such as voltage.

Other examples of this kind of account include \citepbjps{Blachowicz:1997vq}, \citepbjps{Kulvicki:2014cv}, and \citepbjps{Peacocke:2019vm}. Blachowicz argues for what he calls the model interpretation: `the function of analog representation is to map or model what it represents' \citepbjps[p. 83]{Blachowicz:1997vq}. In line with the covariation account, Blachowicz argues that continuity is inessential to analog representation. However, Blachowicz is primarily concerned with examples of analog perception and thought, going beyond the kind of analog representation characterized here. Kulvicki agrees that important examples of analog representation need not be continuous, and that the preservation of structure between what is represented and what is doing the representing is the essential feature. Kulvicki is also focused on more complex examples, including the ways in which these examples support particular kinds of psychological tasks or capacities. Peacocke's view aligns almost perfectly with the Lewis-Maley view to be developed below, although Peacocke's overall purpose is to develop an account of the metaphysics that explains our perceptual capacities (Peacocke's view of digital representation, however is at odds with the Lewis-Maley view).

\citetbjps{Maley:2011jv} builds on Lewis's account by arguing that the notion of primitive or almost primitive physical magnitude is too restrictive, and by offering a more precise characterization of the kind of covariation involved in analog representation. While in one sense Maley's characterization is the narrowest, it is also the most precise. However, we can improve upon Maley's original view, resulting in the following characterization (with the formalism to be explained below):

A representation $R$ of a quantity $Q$ is analog (with resolution $r$) iff:
\begin{itemize}
	\item[1.] there is some property $P$  (the representational property) of $R$ such that the physical quantity or amount of $P$ specifies $Q$; and
	\item[2.] the quantity or amount of $P$ is a monotonic function $f$ of $Q$, and that function is a homomorphism from $Q$ to $P$. Furthermore, let $P_1$ and $P_2$ be values of $P$ that represent quantities $Q_1$ and $Q_2$, respectively. If $|P_1 - P_2| \geq r$, then (without loss of generality) stipulate that $P_1 < P_2$ (that is, let $P_1$ be the smaller of the two). In the case where $f$ is monotonically increasing (non-decreasing), then $Q_1 < Q_2$; if $f$ is monotonically decreasing (non-increasing), then $Q_1 > Q_2$. However, $Q_1 \leq Q_2$ only implies $P_1 \leq P_2$ for monotonically increasing $f$, or $P_1 \geq P_2$ for monotonically decreasing $f$.
\end{itemize}

A few points about this characterization should be highlighted. First, this is not an account or theory of representation in general; rather, this is a characterization of what makes something already taken to be a representation a specifically analog representation. Second, and relatedly, `specifies' in the first clause refers to what it is about the representation that does the representing, and not about how the physical properties of the representation are caused to have the value that they do. Finally, specifying that the function is a monotonic homomorphism allows for maximal generality with respect to which functions will count as preserving an analog relationship. In general, this means that an increase in the property doing the representing necessarily means that what is represented has increased, but an increase in what is to be represented does not necessarily mean that the property doing the representing increases (although it may).

Consider a mercury thermometer. This counts as analog because an increase (or decrease) in a particular property of the thermometer---the height of the mercury---represents an increase (or decrease) in the ambient temperature. Another example is a vinyl record: these are analog because an increase in the `frequency' of the ridges within a groove represents an increase in the frequency of the sound represented (and similarly for the height of the ridges and the amplitude of the sound). Still another is an hourglass: the amount of substance (sand or liquid) in the bottom of the glass increases as the amount of time has elapsed since the glass was turned over.

Importantly, this characterization also counts these examples as analog if they happen to only vary in discrete steps. Consider the hourglass, for example. Many hourglasses use discrete particles, such as sand, whereas others use liquids (which are presumably continuous). On the account offered here, it is the fact that the amount of substance is doing the representing that makes the hourglass analog, rather than whether that substance is continuous or discrete.\footnote{I take it to be a virtue of this account, relative to continuity-based accounts, that we can set aside issues about whether anything is really continuous (properties like voltage, liquids, spacetime itself) in order to determine whether anything is really analog. Such questions simply do not matter for the present account.}

Similarly, this characterization counts analog clocks correctly, whether they vary continuously or discretely. Consider the second hand of an analog clock. On the continuous conception, an analog clock with a hand that sweeps continuously would be analog, whereas one that ticks would not. But on the account here, it is the fact that an increase in the angle of the second hand represents an increase in time that matters, regardless of whether that increase happens continuously or discretely. This also illustrates the relevance of the homomorphism constraint (where an isomorphism would be too strong). Assume that time is continuous, and consider its representation by a discretely-ticking second hand. If the second hand has moved, we know that time has passed. On the other hand, small amounts of time---smaller than the resolution term $r$, which in this case would be one second---can pass without the second hand moving.

The resolution term $r$ also captures the fact that in a discrete analog representation, there may also be small amounts of `jitter' or noise. When the second hand ticks to a new location, it might oscillate very slightly. However, we do not want these oscillations to count as differences in what is being represented. Thus, if the difference in magnitude between two positions is not greater than $r$, then those magnitudes do not represent different quantities.

Finally, some analog representations may represent in a kind of inverse way. Perhaps instead of a thermometer with a substance whose height increases with temperature, one might have one where the height decreases as temperature increases. Also, some analog representations might be seen as increasing in one way, but decreasing in another. As the second hand of a clock increases in angle from 30 to 36 degrees (with respect to the usual `12' at the top), the reflex angle is decreasing from 330 to 324 degrees (a necessary consequence of angular rotation). Roughly speaking, the relations still hold between what is represented and what is doing the represented, except in reverse. However, the inclusion of both monotonically increasing and decreasing functions in the second clause captures such cases.



For present purposes, I will adopt the improved Lewis-Maley account, simply because it is best suited for characterizing the kind of analog representation relevant to understanding analog computation. As mentioned above, this account is relatively limited. However, it is worth mentioning that this account may well be generalizable to coincide with other mirroring accounts, as well as what have been called structural representations \citepbjps{Ramsey:2007wn}, \citepbjps{Shea:2014tq}, \citepbjps{Nirshberg:2020gs}. The Lewis-Maley account as presented covers what we can call one-dimensional representations: one property of a representation represents some quantity (for example, the height of liquid in a thermometer represents the temperature at a single point in space), and it does so by monotonically covarying with that quantity (that is, as the temperature literally increases/decreases over time, the height of the liquid literally increases/decreases over time). In future work, however, this account might be extendable to multiple dimensions to cover analog or structural representations mentioned in other accounts, such as photographs and scale models. For the present essay, however, we only need a precise account of the one-dimensional representations used by analog computers.

\subsection{What makes it `analog'} 
\label{sub:making_sense_of_analog_computation}
Different accounts of analog representation have been based on particular examples, often hypothetical. As mentioned above, where analog computation has been mentioned in this literature, it is usually taken as nothing more than computation using continuous elements, and thus used to support the view that to be analog is to be continuous. This is simply incorrect, as our historical counterexamples have shown. Nevertheless, I endorse the underlying assumption that (all else being equal) we ought to prefer an account of the analog that applies to b analog representation and analog computation.

We saw in \cref{sub:discontinuous_analog_elements} that analog computation is not simply about continuity. Discontinuous analog components were not uncommon in analog computers, and worked alongside continuous elements. This rules out accounts that take analog representation to involve continuity essentially. At the same time, this rules in covariation-based accounts, particularly the Lewis-Maley version. Let us see why.

Consider again the example of the electromechanical step-function generator depicted in Figure \ref{fig:commutator}. Like purely electronic analog computer elements, this device uses voltage to represent a quantity of interest; unlike some other elements, it can only represent a finite set of values that vary discontinuously. The way it represents a quantity is via covariation between that quantity and the voltage: four volts represents a quantity of four; two volts represents a quantity of two.\footnote{Recall once again that this is an exegetical simplification: depending on the particulars of the problem and the machine, the particular values represented by the voltages within a machine may have been uniformly scaled.} This is precisely the Lewis-Maley account of analog representation. At the same time, the Lewis-Maley account also covers continuous elements, such as the ones described in \cref{sub:mechanical_analog_computers} and \cref{sub:electronic_analog_computers}. The monotonic covariation required by that account holds whether the representation varies continuously or discretely.

Now, in order to avoid confusion, it is important to note what the Lewis-Maley account of analog representation does and does not require of analog representation. The idea of monotonic covariation might lead one to think that the example from Figure \ref{fig:commutator} rules out this account; after all, the step function shown is absolutely not monotonic! However, this observation is beside the point. What is monotonic is the relationship between what is doing the representing (the voltage) and what it represents (the values of the function). Consider the first line segment (at the top left) of this function, let us say it is at 9.6 volts. That represents the number 9.6. Next, it steps down to 7.5 volts. The voltage, which is what does the representing, has literally decreased. But so has what is being represented: 7.5 volts represents the number 7.5, and 7.5 is less than 9.6. Similarly for when it steps down again to 2.1 volts: both the voltage and what the voltage represents have literally decreased. In the next jump, up to 5.9 volts, both the voltage and the value being represented increase.

Let us further clarify this account of analog representation by contrasting it with a case of digital representation in a digital computer. Like electronic and electromechanical analog computers, digital computers also use voltage to represent numbers. How digital computers do so is quite different. In short, analog computers represent numbers by representing their quantity, while digital computers represent numbers by representing their names. Specifically, digital computers use a base-2, or binary, representation, which means that a number such as nine is represented as 1001. The way to interpret this sequence is as follows: starting from the rightmost place, there is a 1 in the ones place, a 0 in the twos place, a 0 in the fours place, and a 1 in the eights place; thus, add one to eight, which is nine.\footnote{This is simply the standard way that we represent numbers, although we typically use base-10, or decimal, rather than binary. This is described in more detail in \citepbjps{Maley:2011jv}} In a digital computer, the 0s are typically represented in a circuit element by zero volts, while the 1s are represented by 5 volts. So to represent the number nine, a sequence of at least four circuit elements would be required. From left to right, the first would be at five volts, the second at zero volts, the third at zero volts, and the last at five volts.

Now imagine increasing the value that we want to represent from nine to ten. In binary, this is 1010. This would result in the following change: the first circuit element would remain at five volts, the second element would remain at zero volts, the third would change from zero to five, and the last would change from five to zero. Note the difference between the changes in voltage here and the changes in voltage in the electronic analog computer: as the value being represented in the digital computer increases, some voltages increase, some decrease, and some stay the same. The way they change reflects a change in the elements of the digital representation of the number: the 1s and 0s have to change in a systematic way to represent the digits of the larger number. However, in the analog computer, the voltage does not represent the elements of a representation (the name, or parts of the name) of the number; rather, the voltage represents the magnitude of the number. 

To put this last point in still a different way, observe that the sequence of symbols 1010 is not larger than the sequence 1001, even though what the sequence 1010 represents is larger than what the sequence 1001 represents. This is true even when we look at how the individual 1s and 0s are physically implemented, in this case as voltage levels. But for the analog case, what represents the number ten (ten volts, possibly scaled by some constant $c$) is physically larger than what represents nine (nine volts, again possibly scaled by the same $c$). Both the physical representation and what is represented increase. This is true for other cases as well, such as the height of the liquid in the thermometer, or the angle of the hand in the watch.

Let us now begin putting everything together. What makes analog computers analog is that they use analog representations, understood according to the Lewis-Maley account just described (and what makes analog computers computers will be discussed next). This characterization covers the well-known continuous examples of analog computation such as the EDA, as well as the not-so-well-known discontinuous examples discussed above. Furthermore, the Lewis-Maley account is not an ad hoc characterization of the analog, custom-built for the purpose of making sense of analog computation. Rather, it is a principled account that also makes sense of examples of analog representation outside of computation (as argued in \citepbjps{Maley:2011jv}).

Importantly, requiring the involvement of representations excludes many non-computational systems from the account (and rightly so). For example, virtually any electrical system with a resistor can be interpreted as multiplying voltages. But because voltages in general are not representations of anything, it is not the case that virtually every electrical system is performing analog computations. The point applies to the mechanical and electromechanical components.


\subsection{What makes it `computation'} 
\label{sub:what_makes_it_emph_computation}

If the above is the right way to understand what is analog about analog computation, what is the right way to understand what is computational about analog computation? The short answer is that analog computation is the mechanistic manipulation of analog representations. We have seen what analog representation is about, so let us look at what mechanistic manipulation is about.

First, let us look at what is required of an account of computation simpliciter. A complete account of computation should (among other things) make sense of what makes digital computation and analog computation similar enough such that both are genuine species of computation, yet different enough to count as separate species. Assuming the orthodox view that computation requires representation, the answer to what makes something analog computation is what was just mentioned: analog computation is computation that uses analog representation. Similarly, digital computation is computation that uses digital representation. When we take seriously the fact that analog representation can be discrete, then we have an account that is both principled and does justice to extant examples of computation, such as the ones we examined above. 

What remains is the question of what kinds of manipulations count. The essential idea---which will have to be further developed in future work---is that computation is the mechanistic manipulation of representations, where `mechanistic' is understood in the sense developed by \citetbjps{Piccinini:2015ut}. Unlike Piccinini's account, however, the positive account I have in mind requires the manipulation of representations. Importantly, the mechanisms doing the manipulating must be sensitive only to the physical properties of the representations that are responsible for representing. Given the Lewis-Maley account of analog representation, we can say that a mechanism must manipulate analog representations qua analog representations in order for them to count as analog computational mechanisms. Finally, the manipulations of those representations are such that the result is itself an analog representation.

For example, in the electronic and electromechanical analog computers discussed above, it is the voltage in the circuits elements does the representing. To count as computational, the mechanism (or mechanisms) that manipulate the representations must do so by manipulating their voltages. This is in contrast to mechanisms that manipulate some other property, such as the temperature, of the circuit elements (as a cooling fan might do). Or in the case of mechanical analog computers, the mechanism must manipulate the position, speed, or angle (as the case may be) of the relevant component. Changing the temperature of disk A in Figure \ref{fig:disk} does not count as a manipulation relevant to computation, because temperature is not the property that does the representing.\footnote{Of course, in some other computers, temperature may well be the property that does the representing.} But changing the displacement and rotation does count, because those are the properties that do the representing.

Requiring the involvement of a mechanism separates analog computation from other systems that use analog representations, but where the representations are manipulated by other means.\footnote{The discussion here is focused on devices that compute, rather than the human activity of computation.} For example, the liquid height of a mercury thermometer is an analog representation of temperature. But the thermometer does not have a mechanism that manipulates the level of liquid; rather, the height changes via a natural process of liquid expansion and contraction due to temperature change. Thus, a mercury thermometer does not count as an analog computer.

Now, before concluding, it is worth pointing out that there will be borderline cases where it is unclear whether a computation is being performed or not. For example, is the playing of a record on a standard record player an instance of analog computation? The vinyl record is clearly an analog representation. We also have a mechanism: the turntable spins at a particular speed, and the needle `reads' the properties of the record that are relevant to its being an analog representation. Similarly, placing a weight on a spring scale may twist a dial, indicating its mass. If these count as computations, then they are very simple computations, and it is not clear how simple a mechanistic manipulation needs to be in order to not count as a computation. However, this question is not unique to the present account: it is a question for every account of computation, which we will not try to answer here.\footnote{For example, consider a digital kitchen timer that can only count down from an input setting. It is not clear whether that counts as a computation or not, and if it does, it is also a very simple computation.}


\section{Questions and Objections} 
\label{sec:questions_and_objections}

Before concluding, in this section I will respond to what seem to me obvious questions and objections.

\subsection{Aren't these just hybrid computers?} 
\label{sub:hybrid_computers}
First, some accounts of digital representation would take issue with the idea that components like the discontinuous step-function generator mentioned in \cref{sub:discontinuous_analog_elements} are analog. For example, Haugeland counts this kind of device as digital, simply because it is discrete: in responding to a nearly identical example presented by \citetbjps{Lewis:1971ug}, Haugeland states `I think it's clearly digital - just as digital as a stack of silver dollars, even when the croupier ``counts" them by height' \citepbjps[218]{haugeland1981analog}. So why not say that, when an otherwise-analog (continuous) computer uses a component like this, it is simply a hybrid analog-digital computer?

This kind of response does not do justice to actual hybrid analog-digital computers. To see why requires another brief digression into digital computation. For simplicity, I will limit the discussion to electronic digital computers.

Recall the discussion in the previous section about how digital representation works. Discrete elements are used to represent the digits of numbers, and the digits of numbers represent the number in the usual place-based way. However, when we attend to the step-function generator, it is clear that although it is discrete and discontinuous, it is not digital in the way described above. Thus, it is a mistake to call it so, as Haugeland does. A digital device requires digital representation, and a hybrid analog-digital computer has components that convert analog representations to digital ones (or vice versa).
\begin{figure}[!htb]
	\begin{center}	
	\includegraphics[width=5in]{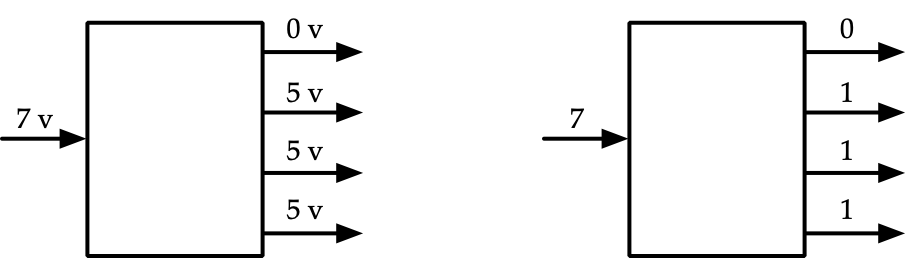}
	\caption{\emph{An analog-digital converter. The physical inputs (voltages) are shown on the left. What those voltages represent is shown on the right. For the analog input, seven volts represents the number seven (although the seven volts may be linearly scaled, per footnote 11); for the digital output, zero volts represents the numeral 0, and five volts represents the numeral 1. These numerals, in turn, represent the number seven when the numerals are interpreted as the digits of a binary representation.}}
	\label{fig:ADC}
	\end{center}
\end{figure}

In short, what makes hybrid analog-digital computers hybrid is not that they use some discrete elements and some continuous ones. Rather, they are hybrid because they use analog components and digital components, where, again, `digital' is understood in the sense discussed above. For example, the interface from an analog to a digital component would include a converter that takes a single voltage level (for example, seven volts) as input and produces as output a series of separate voltages, which themselves represent the binary representation of seven, or 0111 (as illustrated in Figure \ref{fig:ADC}). The conversion from a digital to an analog representation would do the opposite \citepbjps{Korn:1964wy}, \citepbjps{Hyndman:1970uu}. The step-function generator does not operate via a digital representation at all, and requires no conversion to operate with other analog components (for example, the output of the step function generator could be used as an input to some part of the EDA in Figure \ref{fig:MechanicalAnalogComputer}).


\subsection{Is this really even computation?} 
\label{sub:is_this_really_even_computation}
Another objection might go something like this. Analog `computers' have very little to do with obvious paradigms of computers, such as digital computers and Turing Machines. After all, the entire branch of mathematics known as the theory of computation is concern almost exclusively with Turing Machines and other abstract automata, none of which operate on analog representations. Thus, we should not take very seriously the idea that what these analog machines do is a genuine form of computation in the first place.

This objection simply ignores the history of computation as it has actually been practiced. Analog computation was the dominant type of computation for several decades before digital computation became efficient and cost-effective enough to replace it. Analog and digital computers were used to solve similar problems in science and engineering, and although they work quite differently, they were seen as two types of the same kind of machine. Numerous textbooks and research monographs were published with `analog computation' or `analog computers' in the title. Furthermore, the hybrid computing machines just mentioned were not considered to be partially a computer (the digital part), and partially something else (the analog part): they were seen as a single computing machine that operated using two different computational paradigms.

Perhaps there is some positive argument that only digital computation should really count as computation. But that argument will have to explain why so many scientists and engineers were wrong to call certain machines computers in the first place. Without a very strong argument to the contrary, we should consider analog computers to be exactly what scientists and engineers took them to be: genuine computers.


\subsection{The Lewis-Maley account is problematic} 
\label{sub:the_analog_digital_distinction_is_problematic}

Yet another objection might focus on the Lewis-Maley account of analog representation adopted here. One problem with this account is that some examples of representation count as both analog and digital. An example is unary notation, where the number four is represented as a series of four strokes, or four 1s: 1111. We have numerals in places, thus it is digital (for further elaboration of what is meant by `digital', see \S4 of \citepbjps{Maley:2011jv}). But we also have a monotonic covariation between a property of the representation (number of strokes) and what is represented. So this seems to be both a digital and a discrete analog representation. Another example seems even worse: digital representations implemented in contemporary computers. Here, each numeral is represented by a voltage, where the amount of voltage (zero volts or five volts) monotonically increases with what is being represented (a 0 or 1). Again, the Lewis-Maley account seems to classify this as both analog and digital. The received view of the analog---equating `analog' with `continuous'---does not have these faults.

While it is true that the first example is both analog and digital, this is a degenerate sense of digital. Unary, or base-1 notation, can only represent integers, unlike notation in any larger base. To see why, consider what the digits of a base-10 and a base-2 representation mean.
\begin{align}
	314_{10} &= (3 \times 10^2) + (1 \times 10^1) + (4 \times 10^0) = 300+10+4 \nonumber \\
	1001_2 &= (1 \times 2^3) + (0 \times 2^2) + (0 \times 2^1) + (1 \times 2^0) = 8+0+0+1 = 9 \nonumber
\end{align}
In general, for a base $b$ and digits $d_n$\ldots$d_2$ $d_1$ $d_0$, the digital representation is:
\begin{align}
	d_n...d_2d_1d_0 &= (d_n \times b^n)+...+ (d_2 \times b^2) + (d_1 \times b^1) + (d_0 \times b^0) \nonumber
\end{align}
Back to the unary notation:
\begin{align}
	1111_1 &= (1 \times 1^3)+(1 \times 1^2) + (1 \times 1^1) + (1 \times 1^0) = 1+1+1+1 \nonumber
\end{align}

So far, so good. However, things change when we use a decimal point. In bases larger than one, we can represent real numbers, because digits to the right of the decimal place represent negative powers. Thus:
\begin{align}
	29.7_{10} &= (2 \times 10^1) + (9 \times 10^0) + (7 \times 10^{-1}) = 20+9+0.7 \nonumber
\end{align}
In unary notation, however, a decimal point is meaningless:
\begin{align}
	11.11_1 &= (1 \times 1^1)+(1 \times 1^0) + (1 \times 1^{-1}) + (1 \times 1^{-2}) = 1+1+1+1 \nonumber
\end{align}
So, while the unary case does fit the definition of digital on the Lewis-Maley account, this is an easy fix. Biting the bullet turns out to be not that bad: it is both analog and digital, but a degenerate sense of digital. This may seem ad hoc, but it is not: \citetbjps{Turing:1938ui} uses base-2 digital representations of numbers in his discussion of computable numbers. Alternatively, one could simply amend the Lewis-Maley account to stipulate that the digits of a digital representation must be in a base larger than one. Again, this is not as ad hoc as it seems: this would bring the characterization of `digital' in line with how it is understood in Turing's original presentation of computable numbers.

The other concern may seem greater. If even contemporary digital computers use representations that are both analog and digital, then something something has gone awry. After all, the entire point of this essay is that there is an important difference between computers that use analog representations and those that use digital representations! Again, however, the solution is straightforward.

In short, some digital representations will contain discrete analog representations as parts (their digits). Consider the example of the digital representation in Figure \ref{fig:ADC}. In each digit of the whole digital representation 0111 (which represents seven), 0 is represented by 0 volts, and 1 is represented by 5 volts. So, each individual digit, considered in isolation, is a discrete analog representation: larger voltages represent a larger value. However, the entire representation is digital: the rightmost 1, in the ones place,  contributes a value of 1 to what is represented; the leftmost 1, in the fours place, contributes a value of 4 to what is represented. Thus, it is not that this is both an analog and a digital representation. Instead, it is a digital representation that itself has parts implemented as discrete analog representations.

There are undoubtedly other objections that I have not anticipated. My hope is that some of the most important and obvious have been addressed in this section.


\section{Concluding thoughts} 
\label{sec:the_relevance_of_analog_computation}
Let us take stock of what we have done. We have seen different examples of analog computers using mechanical, electronic, and electromechanical mechanisms. We saw that the received view---that analog computation is essentially continuous---fails: some analog computers use discrete, non-continuous elements. We then surveyed some accounts of analog representation, and saw that the the Lewis-Maley account is uniquely able to capture what is analog about analog computation. Next, we briefly looked at how this account of analog computation fits in with more general accounts of computation. Finally, we looked at how to answer some anticipated questions and address some possible objections.

The upshot is that we have a principled account of analog computation that does justice to analog computers implemented in different physical media, using both continuous and discrete representations.

So after all that trouble, why care about analog computation? First, it has simply been neglected: the philosophical attention paid to computation has been almost exclusively aimed at digital computation. As it turns out, analog computation is both interesting in its own right---as well as importantly different from digital computation---in ways that have not been appreciated. Any complete philosophical treatment of computation simpliciter will have to attend to analog computation alongside digital computation.

Second, understanding analog computation offers richer opportunities for the kinds of computation that could play a role in the explanation of the mind and brain. Computationalism about the mind---the view that the mind is literally a computer---has been a major philosophical view for several decades \citepbjps{Piccinini:2009ty}. Computationalism about the brain---the view that the brain is literally a computer---is taken seriously in the sciences of the mind, particularly neuroscience \citepbjps{Shagrir:2006hx}. By couching analog computation in mechanistic terms, this account is applicable to natural (and not just artificial) computational processes like those that may be found in the mind/brain. After all, it seems that some natural processes are mechanistic (\citepbjps{Machamer:2000p1}, \citepbjps{Glennan:vm}, \citepbjps{Woodward:2002dg}). Thus, natural analog computation is a straightforward matter. Some work, such as \citetbjps{Maley:2018us}, has already argued that some neural processes may well be analog---but not digital---computations. Furthermore, many psychologists have already appealed to the analog nature of certain mental processes (seminal examples include \citetbjps{Shepard:1971uf} and \citetbjps{Kosslyn:1994tv}), but without a clear and precise notion of how analog representation and computation might go together. The account here may help provide just such an account. We would be foolish to limit our view of the kind of computation available in these discussions to digital computation alone, given that there is an entirely separate, well-established second kind of computation that has received so little attention.

Finally, this account paves the way for investigations into the possibility of still other types of computation. As sketched above, one natural way to build a general account of computation couches computation in terms of the mechanistic manipulation of representations. Computations are then typed by the kinds of representation involved (which in turn constrains the kind of manipulations possible for the relevant mechanisms). By rejecting the thesis that `analog' and `digital' are jointly-exhaustive opposites, we allow for the possibility of computation that is neither digital nor analog. Specifically, if there are representational types that are neither analog nor digital, and we can specify how such representations are mechanistically manipulated, we can come up with principled accounts of new kinds of computation. This is, of course, largely speculative, but scientists are actively exploring a menagerie of seemingly-exotic computations, including quantum, optical, molecular, membrane, and even physarum (slime mold) computation. It may well be that some of these kinds of computation---if indeed they are species of computation at all---are best characterized as something other than analog or digital computation. We should be in a position to characterize computation more generally, and one first step is a full understanding of what computation could be if it is not digital. I hope to have taken that first step.


\section*{Funding}

This material is based upon work supported by the National Science Foundation under Grant No. 1754974. Any opinions, findings, and conclusions or recommendations expressed in this material are those of the author and do not necessarily reflect the views of the National Science Foundation.

\section*{Acknowledgements}

I would like to thank the following lovely people for their helpful comments and suggestions: Zed Adams, Cameron Buckner, Zoe Drayson, Daniel Estrada, Gualtiero Piccinini, Daniel Schneider, and especially Sarah Robins; the Philosophy Departments at the University of Kansas, the University of Wisconsin, and Western University; audiences at the 2019 SLAPSA, 2018 PSA, and 2018 SSPP conferences; and the MCMP--Western Ontario Workshop on Computation in Scientific Theory and Practice. Finally, the referees for this article provided valuable comments and suggestions for improvement.

\begin{flushright}
\emph{
  Corey J. Maley\\
  University of Kansas\\
  3075 Wescoe Hall\\
  Lawrence, KS 66045\\
  cmaley@ku.edu
}
\end{flushright}

\bibliographystyle{bjps}

\end{document}